\documentclass[
  journal=pasa,
  manuscript=preprint-article,
  year=2025,
]{cup-journal}

\usepackage{amsmath}
\usepackage{amssymb}
\usepackage{hyperref}
\usepackage[normalem]{ulem}
\usepackage{soul}
\hypersetup{
colorlinks=true,
linkcolor=blue,
filecolor=blue,
citecolor =blue,
}
\usepackage[nopatch]{microtype}
\usepackage{booktabs}

\usepackage{tabularx,ragged2e}
\newcolumntype{Y}{>{\RaggedRight\arraybackslash}X}

\title{findAbar: how astronomers may perceive the bar in galaxies differently}

\author{Elizabeth J. Iles}
    \affiliation{Sydney Institute for Astronomy, School of Physics, A28, The University of Sydney, Sydney, NSW 2006, Australia}
    \alsoaffiliation{School of Physics, University of New South Wales, Sydney, NSW 2052, Australia}
    \email[Elizabeth J. Iles]{elizabeth.iles@sydney.edu.au}
\author{Joss Bland-Hawthorn}
    \affiliation{Sydney Institute for Astronomy, School of Physics, A28, The University of Sydney, Sydney, NSW 2006, Australia}
\author{Courtney Crawford}
    \affiliation{Sydney Institute for Astronomy, School of Physics, A28, The University of Sydney, Sydney, NSW 2006, Australia}
\author{Scott Croom}
    \affiliation{Sydney Institute for Astronomy, School of Physics, A28, The University of Sydney, Sydney, NSW 2006, Australia}
\author{Hillary Davis}
    \affiliation{Sydney Institute for Astronomy, School of Physics, A28, The University of Sydney, Sydney, NSW 2006, Australia}
\author{May Gade Pedersen}
    \affiliation{Sydney Institute for Astronomy, School of Physics, A28, The University of Sydney, Sydney, NSW 2006, Australia}
\author{Anne Green}
    \affiliation{Sydney Institute for Astronomy, School of Physics, A28, The University of Sydney, Sydney, NSW 2006, Australia}
\author{Madusha Gunawardhana}
    \affiliation{Sydney Institute for Astronomy, School of Physics, A28, The University of Sydney, Sydney, NSW 2006, Australia}
\author{Miguel Icaza-Lizaola}
    \affiliation{Korea Astronomy and Space Science Institute, 776 Daedeok-daero, Yuseong-gu, Daejeon 34055, Republic of Korea.}
    \alsoaffiliation{Centro de Investigaci\'{o}n Avanzada en F\'isica Fundamental (CIAFF), Facultad de Ciencias, Universidad Aut\'{o}noma de Madrid, 28049 Madrid, Spain\label{inst2}}
\author{Helen Johnston}
    \affiliation{Sydney Institute for Astronomy, School of Physics, A28, The University of Sydney, Sydney, NSW 2006, Australia}
\author{Emily F. Kerrison}
    \affiliation{Sydney Institute for Astronomy, School of Physics, A28, The University of Sydney, Sydney, NSW 2006, Australia}
    \alsoaffiliation{Australian Telescope National Facility, CSIRO Astronomy and Space Science, PO Box 76, Epping, NSW 1710, Australia}
    \alsoaffiliation{ARC Centre of Excellence for All-Sky Astrophysics in 3 Dimensions (ASTRO 3D), Australia}
\author{Yifan Mai}
    \affiliation{Sydney Institute for Astronomy, School of Physics, A28, The University of Sydney, Sydney, NSW 2006, Australia}
    \alsoaffiliation{ARC Centre of Excellence for All-Sky Astrophysics in 3 Dimensions (ASTRO 3D), Australia}
\author{Benjamin T. Montet}
    \affiliation{School of Physics, University of New South Wales, Sydney, NSW 2052, Australia}
\author{Kovi Rose}
    \affiliation{Sydney Institute for Astronomy, School of Physics, A28, The University of Sydney, Sydney, NSW 2006, Australia}
    \alsoaffiliation{Australian Telescope National Facility, CSIRO Astronomy and Space Science, PO Box 76, Epping, NSW 1710, Australia}
\author{Tomas Rutherford}
    \affiliation{Sydney Institute for Astronomy, School of Physics, A28, The University of Sydney, Sydney, NSW 2006, Australia}
    \alsoaffiliation{ARC Centre of Excellence for All-Sky Astrophysics in 3 Dimensions (ASTRO 3D), Australia}
\author{Manasvee Saraf}
    \affiliation{International Centre for Radio Astronomy Research, The University of Western Australia, Crawley, WA 6009, Australia.}
    \alsoaffiliation{Australia Telescope National Facility, CSIRO, Space and Astronomy, P.O. Box 1130, Bentley, WA 6102, Australia}
    \alsoaffiliation{ARC Centre of Excellence for All-Sky Astrophysics in 3 Dimensions (ASTRO 3D), Australia}
\author{Ellen L. Sirks}
    \affiliation{School of Physics, A28, The University of Sydney, Sydney, NSW 2006, Australia}
    \alsoaffiliation{ARC Centre of Excellence for Dark Matter Particle Physics, Australia}
    \alsoaffiliation{Institute for Computational Cosmology, Department of Physics, Durham University, South Road, Durham DH13LE, UK}
\author{Eckhart Spalding}
    \affiliation{Sydney Institute for Astronomy, School of Physics, A28, The University of Sydney, Sydney, NSW 2006, Australia}
    \alsoaffiliation{ETH Zurich, Institute for Particle Physics and Astrophysics, Wolfgang-Pauli-Strasse 27, 8093 Zurich, Switzerland}
\author{Sujeeporn Tuntipong}
    \affiliation{Sydney Institute for Astronomy, School of Physics, A28, The University of Sydney, Sydney, NSW 2006, Australia}
    \alsoaffiliation{ARC Centre of Excellence for All-Sky Astrophysics in 3 Dimensions (ASTRO 3D), Australia}
\author{Jesse van de Sande}
    \affiliation{School of Physics, University of New South Wales, Sydney, NSW 2052, Australia}
\author{Pavadol Yamsiri}
    \affiliation{Sydney Institute for Astronomy, School of Physics, A28, The University of Sydney, Sydney, NSW 2006, Australia}
    \alsoaffiliation{ARC Centre of Excellence for All-Sky Astrophysics in 3 Dimensions (ASTRO 3D), Australia}


\begin{document}

\begin{abstract} 
Bars are ubiquitous morphological features in the observed distribution of galaxies. There are similarly many methods for classifying these features and, without a strict theoretical definition or common standard practice, this is often left to circumstance. So, we were concerned whether astronomers even agree on the bar which they perceive in a given galaxy and whether this could impact perceived scientific results. As an elementary test, we twenty-one astronomers with varied experience in studying resolved galaxies and circumstances, have each assessed 200 galaxy images, spanning the early phase of bar evolution in two different barred galaxy simulations. We find variations exist within the classification of all the standard bar parameters assessed: bar length, axis-ratio, pitch-angle and even whether a bar is present at all. If this is indicative of the wider community, it has implications for interpreting morphological trends, such as bar-end effects. Furthermore, we find that it is surprisingly not expertise but gender, followed by career stage, which gives rise to the largest discrepancies in the reported bar parameters. Currently, automation does not seem to be a viable solution, with bar classifications from two automated bar-finding algorithms tested and failing to find bars in snapshots where most astronomers agree a bar must exist. Increasing dependence on machine learning or crowdsourcing with a training dataset can only serve to obfuscate any existing biases if these originate from the specific astronomer producing the training material. On the strength of this small sample, we encourage an interim best practice to reduce the impact of any possible classification bias and set goals for the community to resolve the issue in the future. 
\end{abstract}

\section{Introduction}

Of all galaxy morphologies observed within the Universe, galaxies with a central bar-like feature account for a significant fraction. Depending on the classification criteria, observational estimates place this fraction of spiral-type galaxies with bars anywhere between $\sim25$--$75$\% of all known galaxies \citep[e.g.][]{schinnerer2002, Aguerri2009, Nair2010, Saha2018}. Bars are also observed in a smaller fraction of lenticular galaxies \citep[e.g.][]{Laurikainen2009}. Bars have long been thought to be more prevalent in denser environments, although this may be simply related to higher interaction frequency in dense environments \cite[e.g.][]{Elmegreen1990, Lee2012, Skibba2012, Pettitt2018, Cavanagh2022}. There is also some debate over whether bars may or may not be more prevalent in early- \citep[e.g.][]{Combes1993, Masters2011, Skibba2012, Vera2016, CervantesSodi2017, Erwin2018} or late-type  \citep[e.g.][]{Erwin2018, Tawfeek2022} spirals. It has also long been established that our Milky Way galaxy is host to a bar in the central region of the galaxy \citep[e.g.][]{Blitz1991, Nakada1991, Paczynski1994, Zhao1994}. As such, this ensures the significance of bars to developing an understanding of both the local Galactic environment, as well as the formation and evolution of galaxies on a cosmological scale. 

Intrinsic properties of bar features are already being used to probe the likely evolutionary histories of their host galaxy. For example, studies have probed how the length of the bar feature may be correlated with whether the galaxy is an early- or late-type spiral, with shorter bars in late-type galaxies \citep[e.g][]{Elmegreen1985, Combes1993, Erwin2005, Aguerri2009, Erwin2019}. At the very least, bar lengths appear to increase with increasingly massive galaxies \citep[e.g.][]{Kormendy1979, Erwin2005, Diaz-Garcia2016}, although observations do not consistently show a trend of increasing bar length with redshift \citep[e.g.][]{Sheth2008, Kim2021}. This has been suggested to relate to the available gas fractions of the host-galaxy, since a higher gas fraction (i.e.\ as seen in most late-type galaxies) is shown to suppress the growth of bars in galaxies \citep[e.g.][]{Bournaud2005, Berentzen2007, Athanassoula2013, Bland-Hawthorn2024}. These features are not only useful for considering the universal trends of galaxy formation and evolution, but are also applicable to the impact that a bar itself may have on many internal processes of the host-galaxy. 

Bars are well known to impact gas flow in the central regions and correspondingly, impact the star forming conditions across the disk. For example, many studies have probed whether the star formation rates (SFR) and efficiencies (SFE) vary across the morphological features in galaxies, including most prominently these central bars, with the bar often having a lower SFE when compared to the disk or spiral arms and also found to be lower than a region called the \textquoteleft bar-ends’ in some galaxies where the bar and arms are intersecting \citep[e.g.][]{Downes1996, Sheth2002, Momose2010, Watanabe2019, Querejeta2021, iles2022}. The impact of bars on the host-galaxy is not limited to gas flow and star formation, but can also be found in studies related to the total angular momentum distribution \citep[e.g.][]{Weinberg1985, Athanassoula2005}; disk-halo and disk-bulge interaction and evolution \citep[e.g.][]{Athanassoula2002, Valenzuela2003, Kormendy2004, Jogee2005, Athanassoula2013, Kruk2018}, Active Galactic Nuclei (AGN) presence and activity \citep[e.g.][]{Shlosman1989, Alonso2014, Garland2023}, as well as the factors preceding the end-phase of galaxy evolution, such as gas depletion and quenching \citep[][]{Masters2012, Spinoso2017, Fraser-McKelvie2020, Geron2021}. 

So, it is important to understand where, when and how bars form, as well as how these features will influence the host galaxy around them. However, to achieve this, we require a precise, practical and consistent method for identifying what exactly comprises the bar feature within various galaxies, both real and simulated, which we currently lack. Consider the most straightforward delimiter of a galactic bar, the bar length, for example. It is common to describe the extent of a bar in a range of ways: the absolute length of the bar, via the bar radius $R_{\rm bar}$ in length units (often in kpc or arcseconds); or, in relative units, such as in units of the radial scale length which has been normalised to the exponential profile; or at least normalised by the radial extent of the host-galaxy. This length can also be determined in a number of ways and many are regularly used in various studies of bars, depending on perhaps the data-type; the availability of resources; and sometimes simply, even the personal preference of the lead researcher. For example, Fourier analysis, usually using primarily the $m=2$ mode of the azimuthal profile, has been used for many years and remains popular \citep[e.g.][]{Elmegreen1985, Ohta1990, Aguerri1998, Garcia-Gomez2017, Pettitt2018}. Alternatively, it has also been common, particularly for observational surveys, to use a range of analytical isophotal ellipse fitting \citep[e.g.][]{Abraham1999, Laine2002, Erwin2005, Gadotti2006, Marinova2007, Menendez-Delmestre2007, Consolandi2016, Jiang2018} or photometric decomposition type methods \citep[e.g.][]{bureau2006,Reese2007, Gadotti2008, Durbala2008, Durbala2009, Weinzirl2009, Kruk2018}. Recently, efforts have been in place to automate the determination of the barred region in galaxies, in particular, making use of essentially non-parametric, deep learning models, via classification, regression and segmentation techniques \citep[e.g.][]{Abraham2018, Cavanagh2020, Fluke2020, Cavanagh2022, Huertas-Company2023, Cavanagh2024}. Of course, the most simple classification method is still by \textquoteleft eye’ but visual classification of large numbers of galaxies can prove to be taxing, especially in the era of large surveys, and it has long been recognised that this process has at least some non-negligible scatter between observers \citep[e.g.][]{Lahav1995}. Despite this, visual classification remains popular, either as the primary method of classification or, at least, as a \textquoteleft sanity check’ for more automated classification methods. Usually, this is completed by astronomers for their own research purposes, limiting the observers making these classifications to the order of the size of a single research group (i.e.$\sim1$--$10$ astronomers). However, with open crowdsourcing initiatives such as GalaxyZoo, it is also possible to complete morphological classifications with a significant statistical sample number, reducing the effect of individual observer differences which has been used for a number of prominent bar-related studies \citep[e.g.][]{Hoyle2011, Masters2011, Masters2012, Skibba2012, Melvin2014, Kruk2018, Masters2021, Geron2021}. However, with so many different methods for classifying a galactic bar in active use, is there any assurance that the bars identified in a given study are consistent with the bars identified in other various studies? Is it possible to directly compare even the simple morphological properties of bars identified by different researchers, at different times, on different data, via different classification methods? That is, are we really \textquoteleft seeing’ the same bar each time? 

This paper is the result of an attempt to assess how we as astronomers, the so-called experts, perceive the \textquoteleft bar’ in galaxies on the scale of a research group in a relatively large astronomy department, and highlight where the existing systematic discrepancies in our classifications could have significant implications on our perceived results. The immediate goal, in the short-term, is broadly quality control; to raise awareness for these issues and, in doing so, strongly encourage all researchers who wish to present work related to galactic bars to be as detailed and specific as possible in describing the bar definition process used. We argue that this articulation of the bar definition in all future publications must, in the interim, become an astronomy-wide best practice before a common definition or method for defining the bar can be found. The subsequent article is structured as follows: in Section~\ref{s:method} we describe the methodology for the assessment of bar features; in Section~\ref{s:results} we demonstrate how visually identified bar properties can differ depending on various personal factors, even within a group of professional astronomers; in Section~\ref{s:significance} we provide an example of how this result may be of scientific significance to the ongoing priorities of the astronomy community; and finally, in Section~\ref{s:solutions} we discuss automation as a possible solution and provide a list of advice for the mitigation of any bias which may be present. The conclusions of this work are presented in Section~\ref{s:conclusion}. 

Finally, we acknowledge that the results presented herein are not meant to be taken as a complete description of all practicing astronomers, but rather these results should serve as a call to action. If these systematic differences appear within the singular research environment of the authors, it is similarly possible that such differences can occur between researchers in similar contexts around the world. For now, there is no way to quantify the global extent of these trends, but we argue that in the way we practice astronomy today it would not be unexpected that a difference in perception between even just two researchers working on a similar problem could lead to different outcomes in our understanding of bars in galaxies. This work is intended to remind the field at large of our inherent differences in perception as experts and, concurrently, to encourage the search for a method that allows complete repeatability in the identification of visual/morphological features in galaxies. 

\section{Method}
\label{s:method}
The focus of this study was not to investigate the inherent trends or biases which may exist within any visual classification schema, but to identify whether we, as astronomers, as experts, may not be comparing the same features in similar studies undertaken by different researchers. Hence, we as members of a single community of astronomers, independently contributed bar definitions for a set of $200$ snapshots comprising of the first 1\,Gyr evolution for two barred galaxy simulations from \citet{iles2022} with the goal of constraining our understanding of how we, as a community, may perceive the bar in these galaxies. 

\begin{figure}[hbt!]
	\includegraphics[width=\columnwidth]{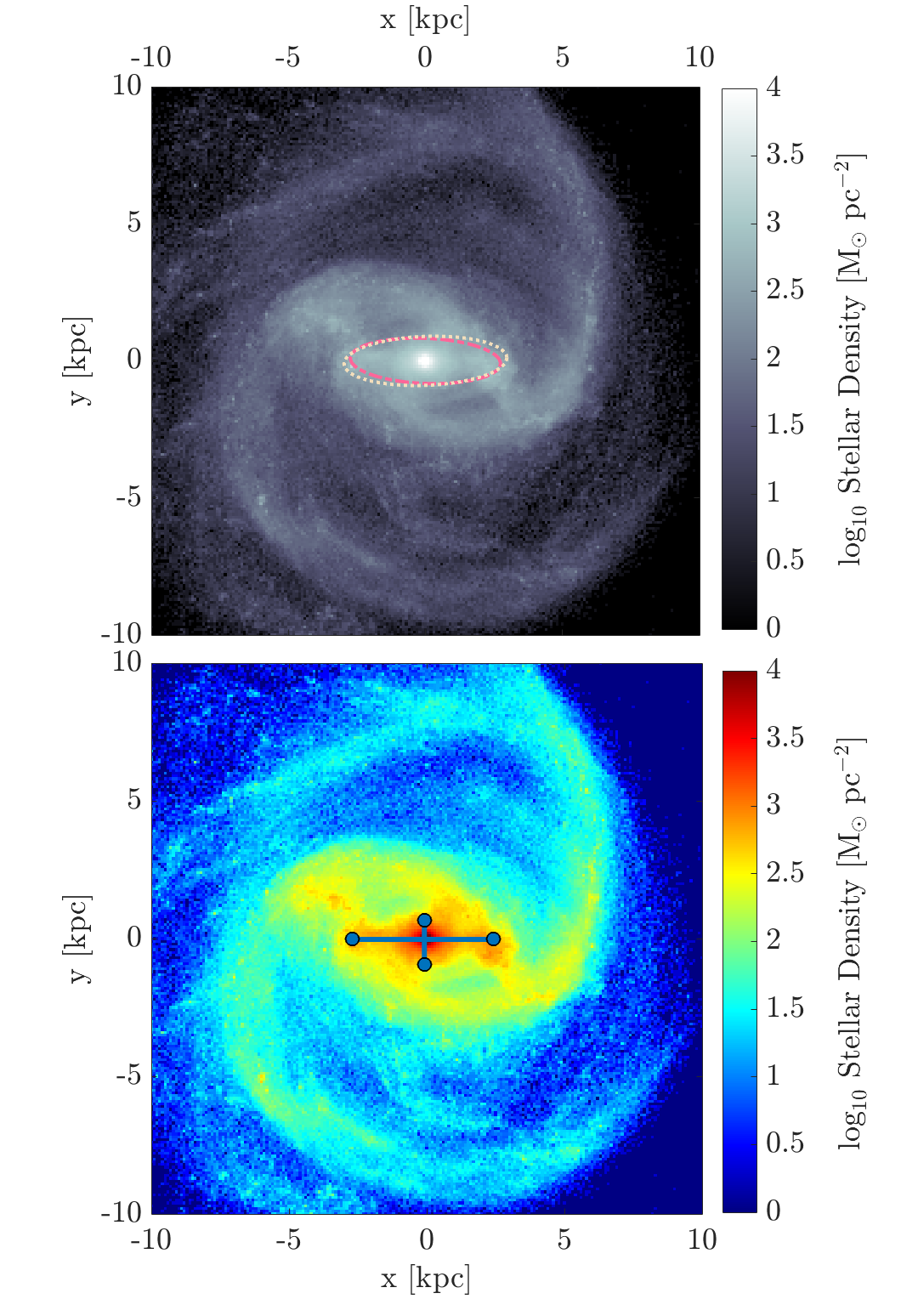}
    \caption{\emph{Upper:} A standard face-on stellar density map for one of the barred galaxies included in \emph{findAbar} \citep[IsoB; ][]{iles2022, iles2024} with coloured ellipses tracing two different responses from the participating astronomers. \emph{Lower:} An example of the image used for \emph{findAbar} and the line selections for classifying the semi-major and semi-minor bar axes.}
    \label{f:matlab_bars}
\end{figure}

\subsection{Details of the Classification Method}
The galaxy images were drawn from a series of $N-$body, Smoothed Particle Hydrodynamics (SPH) simulations providing a relatively large sample of similar but not identical snapshots of bar formation and evolution. These simulations, evolved via \textsc{Gasoline2} \citep{Wadsley2017} and initially tailored to the well-observed, nearby barred-galaxies \textsc{NGC\,4303} (IsoB; isolated bar formation scenario) and \textsc{NGC\,3627} (TideB; bar formation affected by tidal interaction), have previously been used to study morphologically dependent trends in star formation \citep{iles2022}, as well as radial migration and disk metallicity gradients \citep{iles2024}. In these simulated galaxies, the bars were originally determined by \citet{iles2022} to form at around $600$ and $400$\,Myr simulation time, respectively, so we anticipate this sample to comprise both the period preceding and including bar formation, to probe how early in the formation process astronomers may visually perceive a bar in the stellar structure, as well as a short subsequent period over which to assess the differences in bar definition post-bar formation, while the bar settles. This sample, while large in number, is limited by containing only early-stage bars, presented in only the stellar component from simulated data, with images in a face-on orientation. However, it forms a simple, straightforward and systematic test case where any variation in the classification should come directly from the astronomer, rather than any data-type familiarity or image quality/resolution effects, for example.

The practical bar identification process required each astronomer to identify the semi-major and semi-minor axis of an ellipse which would encapsulate any perceived bar in an image for each snapshot of these simulated galaxies. This was achieved by manually drawing a line for each axis on a projection of the face-on stellar density, which had the colour map and contour levels fixed. In practice, for such line-drawing, we took advantage of \textsc{matlab}’s regions of interest (ROI) in the \textquoteleft image processing’ toolbox \citep{matlab}. The end-points of these lines were averaged to produce the final values for bar length and bar width. The pitch angle could also be subsequently calculated with simple geometry. There was only one opportunity for each astronomer to assess each stellar density image and each image was sequential in the evolution of the galaxy. We called this process \emph{findAbar}. For reference, an example of the bar classification image is included in Figure~\ref{f:matlab_bars} with the regular image of the galaxy \citep[e.g.][]{iles2022} placed above (Figure~\ref{f:matlab_bars}) overlaid by two randomly selected ellipses from the \emph{findAbar} responses to demonstrate the bars identified. 

There are naturally also limitations to this method. For example, the sequential evolution of the simulated galaxy images assessed could permit the classifying astronomer to predict rather than observe structures in the image. A more comprehensive study would perhaps require the astronomer to instead assess the same barred snapshot at different, random or multiple times throughout the bar finding process ($\mathcal{O}$[minutes]) and, in fact, at different stages in their work with barred galaxies ($\mathcal{O}$[years]). However, both these additional requirements would also require a larger time commitment than we felt was necessary for this kind of elementary test. Additionally, it is well established that colour may affect the way the human eye is able to perceive a given feature in an image \citep[e.g.][]{Crameri2020,Bosten2022,Hiramatsu2023}. If this method were to be used to identify bars in these galaxies with the intent to apply these classifications for some scientific investigation, we would need to ensure that the specific bar parameters measured in this way were not strongly biased by our colour map and colour limit choices. This would again necessitate the introduction of more images (and more time) for assessment, varying both the colour map and limits on the same snapshot for multiple assessments. However, these limitations do not effect the comparative power of these results. In this case, each of the contributing authors had the same experience and so we argue that, while there must inherently be an impact from these choices, the whole set of classifications should be subject to the same effects and, therefore, should remain comparable as a simple assessment for our goal to determine whether the bars we, as a community of astronomers, are classifying in the same galaxies are, necessarily, the same bars.

\subsection{A Note on Astronomer Demographics}
\label{ss:piecharts}

Initially, the contributing authors each volunteered their classifications somewhat organically, through word-of-mouth discussion about the project leading to those interested taking the opportunity to participate. However, as differences in responses did indeed begin to appear, we began to wonder about the cause of these differences. As such, we have attempted to classify the participating astronomers by three attributes: experience with studies of resolved galaxies (i.e.\ where bars and other morphological features are evident); career level; and gender. Together, we have gathered a sample of professionals who are relatively evenly distributed in each of these areas in an attempt to categorise whether the most significant differences in classification could be systematic, with some inherent cause.  

\begin{figure}[hbt!]
	\includegraphics[width=\columnwidth]{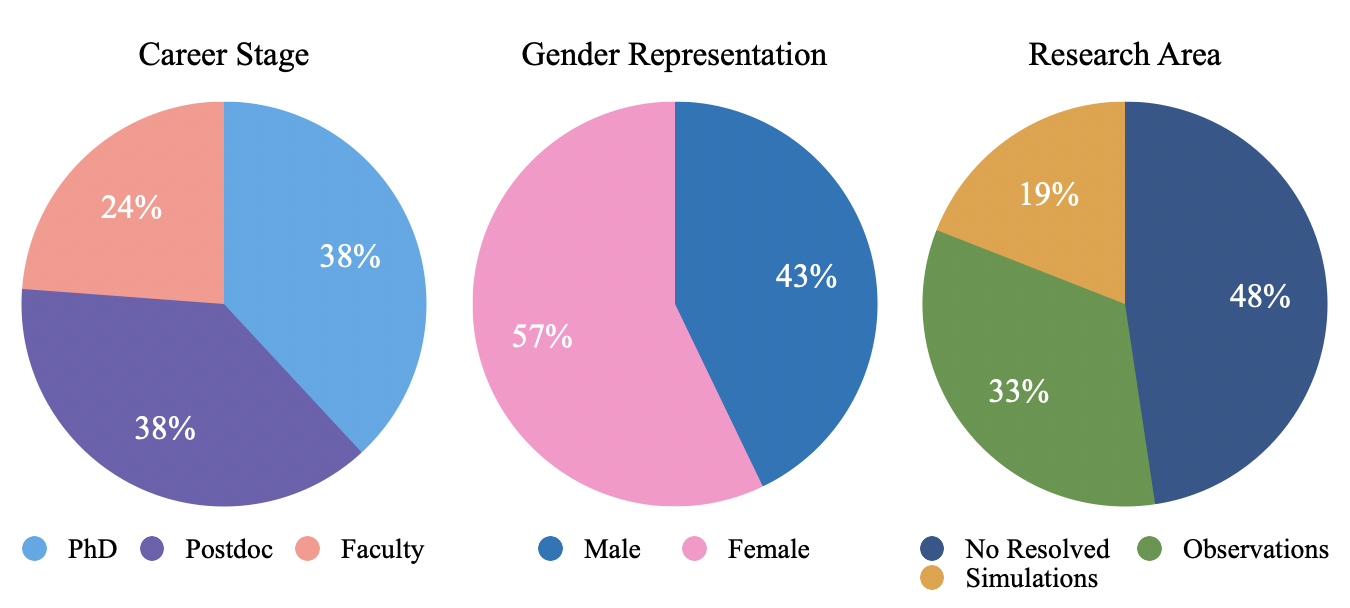}
    \caption{The distribution of participants by career stage, gender and experience working with galaxies displaying bar structure.}
    \label{f:piechart}
\end{figure}

\begin{figure}[hbt!]
	\includegraphics[width=\columnwidth]{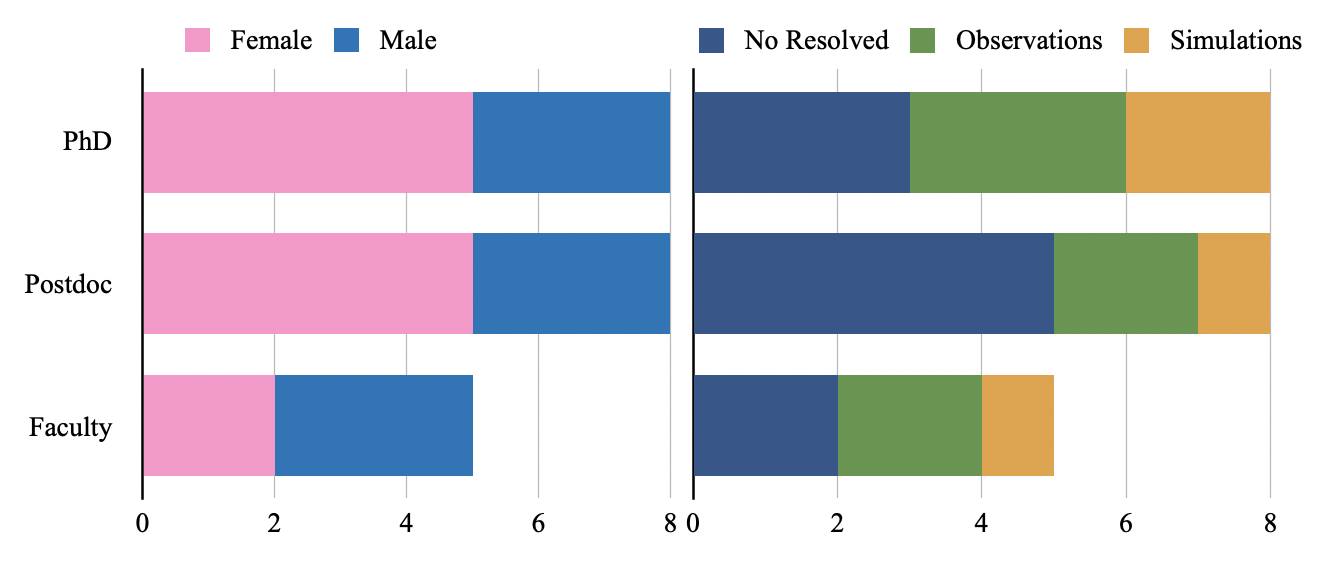}
    \caption{A subdivision of the distribution of participants indicating how the three participant attributes are distributed within the context of career stage.}
    \label{f:barchart_postype}
\end{figure}

Figure~\ref{f:piechart} demonstrates the distribution of participating astronomers by career stage (PhD students - \emph{PhD}, Post-doctoral researchers - \emph{Postdoc} and senior staff with continuing positions - \emph{Faculty}, are labels used as a metric for academic age), gender (\emph{Male}, \emph{Female}, initially not considered but introduced due to the emergence of possible trends) and experience working with galaxies in which structures, such as the bar and spiral arms, may be resolved (astronomers who do not work with galaxies or resolved galaxy morphology - \emph{No Resolved}, observational astronomers - \emph{Observations}, simulation-focused/theoretical astronomers - \emph{Simulations}, initially considered the most likely attribute for variance). Additionally, in order to validate that no singular attribute should dominate all other attribute groups, we also include Figure~\ref{f:barchart_postype}. In this figure, the different attributes are displayed relative to academic level. This is a visual demonstration that the distribution within each of these attributes is consistent with the general distribution presented in Figure~\ref{f:piechart}.

In the interest of effectively using observational and simulation results seamlessly together for scientific advancement in the field, it should be necessary to confirm that the bars (or barred regions of galaxies) defined by those working with simulations are consistent with bars defined by those working primarily with observations. It is with this intention that we have aimed to include an equal number of simulators and observers of barred galaxies, however, we have been limited by a non-portable classification method requiring physical presence and membership in our community, so simulators are slightly under-represented by circumstance of our department demographics. We additionally compare the classification of bars from astronomers familiar with barred galaxies, via their own research, with astronomers who do not work with these objects in their daily research. These astronomers are familiar with astronomical analysis techniques and galaxy evolution theory, but not day-to-day practice, and therefore serve as a control group for any biases or trends that may arise between the simulation and observation groups. It follows then, that the number of contributing astronomers who have experience studying resolved galaxies should be approximately equal to those with no specific experience in this area for a balanced comparison. We note that not all researchers of bars and barred galaxies necessary spend time classifying these features themselves, and not all those who study bars do so in purely the stellar component of these galaxies, however, as this is a self-reflection and elementary assessment, we believe that this general distinction between our expertise is reasonable. 

That we have divided career stage into these broad categories of PhD students, post-doctoral researchers and faculty/staff as a metric for academic age (and thus, time spent in the field), is also a rather loose classification as any given astronomer may not have had the same research focus for the length of their career and so, cross-correlation with expertise is tenuous. However, this is a very reasonable metric for how long it has been since any given astronomer first encountered the definition of bar-like morphology in galaxies and how long they may have been exposed to ongoing discourse within the field. Additionally, we do not include students before the PhD stage because these students will not necessarily have had sufficient time to be exposed to the field and develop opinions or biases. Ideally, the career stage of all participating astronomers would be equally split between the three levels, but the reality is that faculty members are the smallest group in the department and usually the most overwhelmed with competing requirements for their time which leads to a slightly lower contribution rate from this group overall. Regardless, it should at least be possible to assess if there is evidence of the bar definition perhaps changing with time in the field.

A difference in bar definition based purely on the gender attribute should be unlikely but studies have long demonstrated gender differences in a range of areas related to learning, cognition and perception \citep[e.g.][]{Heidl2013,Qian2022, Khaleghimoghaddam2024, Rey-Guerra2024}. Astronomy, like most STEM professions, is a traditionally male-dominated field which has recently moved to improve gender balance, equity and diversity but remains historically unequal in numbers of publications and/or senior positions \citep[see e.g.][]{Stevenson2025,Tran2025}. Despite the relatively small sample group, we find that our participating astronomers are mostly evenly distributed in gender, with a slight skew towards those identifying as female. This is, perhaps surprisingly, indicative of our community demographics, where much work has been done at an institutional level to reach (almost) gender-parity. It is for this reason that we also consider the gender attribute of our perspective on bars. We also note that while the gender attribute in this work appears to be binary, the authors wish to acknowledge that this may not necessarily be the case within a different sample of astronomers currently working in the field.

Finally, we were somewhat concerned that if some personal attribute, such as these, could be associated with a significant bias in the bar classification process, whether that attribute may dominate and falsely imply a similar bias correlated with another attribute. As demonstrated by Figure~\ref{f:barchart_postype}, we have done our best within our limited circumstances to reduce the likelihood of this occurring. All the groups of attributes that we compare are relatively balanced with respect to all other groups of attributes. However, in a larger study with significantly higher participation numbers, we suggest that it could be significant to subdivide the responses by interrelated attributes (e.g.\ compare the classification of male simulators and female simulators against similar observers). Additionally, these three attributes were selected here as the most obvious, first test cases. Other attributes may also be significant which have not been assessed here. For example, we have considered that cultural background or the location/research environment where someone first studied astronomy or completed a PhD may be an important influence on their future preconceived ideas and academic perceptions. However, we are limited by our community size and demographics, so have been unable to test these and similar attributes, as we are already working with small number statistics. We hope that the evidence presented here will be motivation for the field at large to undertake such a large and comprehensive study in the future. 

The responses presented herein were recorded over the period between January-April 2024.

\section{Results}
\label{s:results}

\subsection{A Short Summary of Astronomer Responses}
The primary attributes of a bar can be considered as the length, width and orientation. Here, we produce an ellipse encompassing the bar with semi-major axis $(R_{\rm bar})$ accounting for the bar length, the semi-minor axis $(b_{\rm bar})$ accounting for the width and the angle $(\phi_{\rm bar})$ accounting for the orientation of the bar relative to Cartesian co-ordinates with the plane of the disk in the $xy-$plane. We also introduce two further response metrics: the Axis Ratio, that is the relative roundness of the bar $({b_{\rm bar}}/{R_{\rm bar}})$ where the two extremes are such that a value of 1 would correspond to a circle and 0 to a straight line; and the percentage of participating astronomers to state that a bar exists in a given galaxy snapshot, with a value of 100\% the case where all astronomers identify a bar exists while 0 indicates no-one believed a bar to exist in that snapshot.

\begin{figure*}[ht!]
	\includegraphics[width=\textwidth]{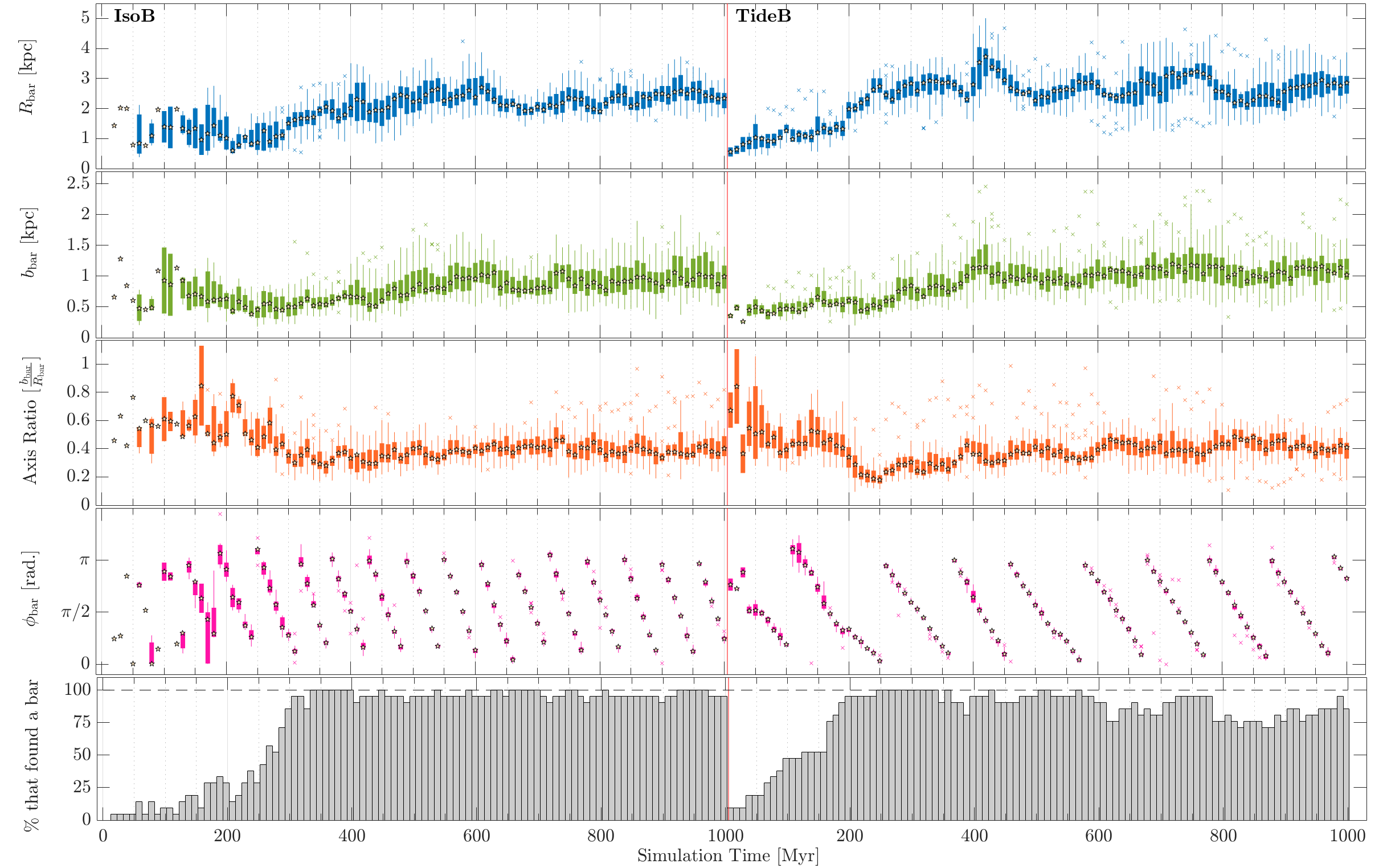}
    \caption{Representation of the full \emph{findAbar} dataset in general parameters: $R_{\rm bar}$, $b_{\rm bar}$, Axis Ratio, $\phi_{\rm bar}$, and fraction of individuals who found a bar at each time-step. The bar parameter responses are presented by a boxplot for each snapshot. The central mark (yellow star) indicates the median response; the box spans from the 25th-75th percentile; whiskers (thin lines) extend to the most extreme data point within $Q_3+1.5(Q_3-Q_1)$ and $Q_1-1.5(Q_3-Q_1)$ and any values more extreme than these limits are classed as outliers (cross marks). The grey bars indicate the percentage of participants who identified a bar in a given snapshot, with the dashed line at 100\% found a bar.}
    \label{f:allbox}
\end{figure*}

The responses for each of these parameters are presented in Figure~\ref{f:allbox} in the form of a box-and-whisker plot at each snapshot of galaxy evolution included in \emph{findAbar}. The median result is identified with a mark on the box (small yellow star). This median value is used as a single value representing the response in these parameters where necessary for the subsequent analysis. The spread of the data is depicted by the solid box and thin line whiskers. Here, the box spans from the 25th to 75th percentile of the responses while the whiskers extend to the most extreme point in either direction not classed as an outlier. Using the basic determination in \textsc{matlab} for box plots, an outlier is considered any point greater than  $Q_3+W(Q_3-Q_1)$ or less than $Q_1+W(Q_3-Q_1)$, where $Q_1$ and $Q_3$ are each the 25th and 75th percentiles, while $W=1.5$ is set as a weighting, such that these whiskers should cover $\pm2.7\sigma$ (or 99.3\%) of values if the data were to be distributed by a normal function \citep{matlab}. Outliers in Figure~\ref{f:allbox} are represented by cross marks at the value.

The median values for each parameter clearly change over the period of evolution in each disk. The easiest to see is the disk rotation as the bar orientation angle ($\phi_{\rm bar}$; pink; panel 4) systematically precesses through the great circle as time passes. However, both the bar length and semi-minor axis lengths ($R_{\rm bar}$; blue; panel 1; and, $b_{\rm bar}$; green; panel 2) can be seen to not only grow in the early stages of evolution, but also to modulate significantly in later periods. Despite the scatter in responses, it is likely that this is a real effect in the bar growth phase, before settling into secular evolution after the relatively short 1\,Gyr period probed here \citep[see e.g.][]{iles2022}. This length variation is most obvious in the $R_{\rm bar}$ measurement for the TideB disk. As this disk is influenced by a flyby interaction with a small stellar companion (i.e.\ a dwarf galaxy) and the bar formed is both stronger and quicker to form than the isolated disk, IsoB. However, interestingly, the Axis Ratio appears similarly stable in both disks post-bar formation, indicating that whatever is triggering the length variation is equally affecting the semi-minor axis (or width) of the bar. Whether this is a true physical effect or, instead, related to a predetermined expectation in our perception of the bar is not currently possible to determine.

\begin{table*}
\centering
\begin{tabular}{|l|l|l|l|l|l|l|l|l|l|l|l|l|} 
\hline
 & \multicolumn{6}{l|}{\textbf{IsoB}} & \multicolumn{6}{l|}{\textbf{TideB}} \\ 
\hline
\begin{tabular}[c]{@{}l@{}}\textbf{Parameter}\\\textbf{(Average)}\end{tabular} & Med. & IQR & Rel. & \begin{tabular}[c]{@{}l@{}}Med.\\($\ge75$\%)\end{tabular} & \begin{tabular}[c]{@{}l@{}}IQR\\($\ge75$\%)\end{tabular} & \begin{tabular}[c]{@{}l@{}}Rel.\\($\ge75$\%)\end{tabular} & Med. & IQR & Rel. & \begin{tabular}[c]{@{}l@{}}Med.\\($\ge75$\%)\end{tabular} & \begin{tabular}[c]{@{}l@{}}IQR\\($\ge75$\%)\end{tabular} & \begin{tabular}[c]{@{}l@{}}Rel.\\($\ge75$\%)\end{tabular} \\ 
\hline
$R_{\rm bar}$ [kpc] & 1.93 & 0.60 & 31.1\% & 2.21 & 0.59  & 26.7\% & 2.37 & 0.56 & 23.6\% & 2.65 & 0.58 & 21.9\%  \\ 
\hline
$b_{\rm bar}$ [kpc] & 0.78  & 0.33 & 42.3\% & 0.82 & 0.34 & 41.5\% & 0.87 & 0.27 & 31.0\% & 0.95 & 0.29 & 30.5\%  \\ 
\hline
Axis Ratio [$b_{\rm bar}/R_{\rm bar}$] & 0.43  & 0.12 & 27.9 & 0.37  & 0.10  & 27.0 & 0.39 & 0.13 & 33.3 & 0.37 & 0.10 & 27.0  \\ 
\hline
$\phi_{\rm bar}$ [rad.] & 1.75 & 0.19 & 10.9 & 1.75 & 0.12 & 6.9 & 1.68 & 0.12   & 7.1  & 1.61 & 0.09 & 5.6   \\
\hline
\end{tabular}
\caption{A single average value determined for each of the standard parameters: $R_{\rm bar}$, $b_{\rm bar}$, Axis Ratio and $\phi_{\rm bar}$ from the median response (Figure~\ref{f:allbox} star-shaped points) in each time period (Med.) and the corresponding IQR (Figure~\ref{f:allbox} boxes) and the relative size of the IQR spread to the Med. value (Rel.). Each simulated disk in the sample (IsoB, TideB) has two sets of values for the total simulation time and the periods where $\ge 75$\% of astronomers agree a bar exists.}
\label{t:stats}
\end{table*}

In addition to the inherent variation of the bar parameters over time, there also appears to be some distinguishable variation in the spread of the responses with certain periods. That is, some time periods appear to have a larger spread in response data (box+whisker size) than others. Other than the earliest periods of each simulation, the bar orientation angle $\phi_{\rm bar}$ appears to have the most agreement (smallest spread of results) between participating astronomers. This makes the bar orientation the most straightforward metric to measure, there is little ambiguity in the angle mapped by this central feature. What little spread there is, is more likely to be related to random variation in the exact pixel chosen by each astronomer at each time period, than any significant variance in the result. Conversely, there are certainly periods where there was strong agreement in the bar length (i.e.\ $R_{\rm bar}$ at $\sim700$\,Myr in IsoB; at $\sim375$\,Myr in TideB) but also considerably less agreement ($\sim$more than double box length) in nearly adjacent periods (i.e.\ $R_{\rm bar}$ at $\sim600$\,Myr in IsoB; at $\sim400$\,Myr in TideB). It is tentatively theorised that this may be related to the arm orientation at the bar edges obscuring where the bar may truly end for some periods and becoming more clear at others. However, whether this is also a random variance effect cannot be completely discounted at this time. We suggest a larger, more complete study which includes systematic measurements of the arm region morphology would be necessary to determine whether such differences in the variance are true features of the bar classification and whether this is related to the bar-arm connection. While such tests have not currently been systematically conducted, visual inspection of the galaxy evolution in both the stellar and gas density maps does not indicate that these periods are special in any obvious way. With future studies, we hope to expand upon this trend and its possible physical or psychological causes.

Finally, the last panel in this figure contains the counts (grey bars) of responses for each time period, indicating the fraction which identified a bar to be present. The dashed black line is at 100\% where all would agree a bar exists. It is perhaps significant that there are relatively few periods where this appears to be the case (25/100 in IsoB, 15/100 in TideB; total=40/200). However, after the initial period of bar formation, where the numbers clearly grow sharply as the bar itself grows clearer, most ($\gtrsim75$\%) contributing astronomers continue to perceive a bar for the duration of the simulation. This can perhaps be attributed to the fact that the simulated snapshots were presented to participants in order of simulation time. Hence, a belief that, once a bar is formed it will continue to be present in the disk for a long period of time, may make us feel that we should continue finding a bar in subsequent snapshots, when perhaps a bar would not be so easily identified were these galaxy snapshots seen in isolation or out of order. We must consider if this is a prevailing belief in astronomy and/or whether there is some unidentified factor in the attributes of those $\sim25$\% of participants who were willing to believe the bar had dispersed at later periods.  

As a summary, we present Table~\ref{t:stats} where we calculate an average value for each of the standard parameters over the duration of the simulation. We use the median of responses (Med.) as the \textquoteleft true’ value determined by our group of astronomers and the interquartile range (IQR) as a measure for the dispersion of these responses at each time period. Additionally, we demonstrate how each of these results changes if we are only to consider the time periods where a significant fraction of astronomers agree the disk is barred. In all cases, the dispersion (IQR) between the responses for each parameter decreases relative to the true value when we consider only periods where $\ge75\%$ of astronomers agree a bar exists. So, this could mean with more astronomers the differences in individual classifications become less significant overall, or that the bar feature becoming more obvious is driving the classifications to become more coherent. We expect it is the latter. Additionally, we can see that the parameter which appears most difficult to classify is the bar semi-minor axis or width. This has the largest IQR compared to the magnitude of the true value and, while this is not surprising, it calls into question the value of this parameter for studies of the bar in galaxies. For the subsequent analysis, we collapse this into the composite parameter of the Axis Ratio. The bar length ($R_{\rm bar}$) and Axis Ratio for both disks have a similar spread of responses accounting for $\sim20$--$30$\% of the true value, while the orientation angle ($\phi_{\rm bar}$) has consistently lower ($\sim5$--$10$\%) as can be seen visibly from Figure~\ref{f:allbox}. However, taking only one value over the simulation is not necessarily realistic for a bar, even on such a relative short evolutionary time period ($\sim 1$\,Gyr), although it is seen in the literature more often than one would like to believe. Regardless, this serves to demonstrate that there is significant and ongoing variation in the classification of all the standard bar parameters which cannot be averaged into a less significant contribution with more time (and more snapshots). 

\begin{figure*}
	\includegraphics[width=\textwidth]{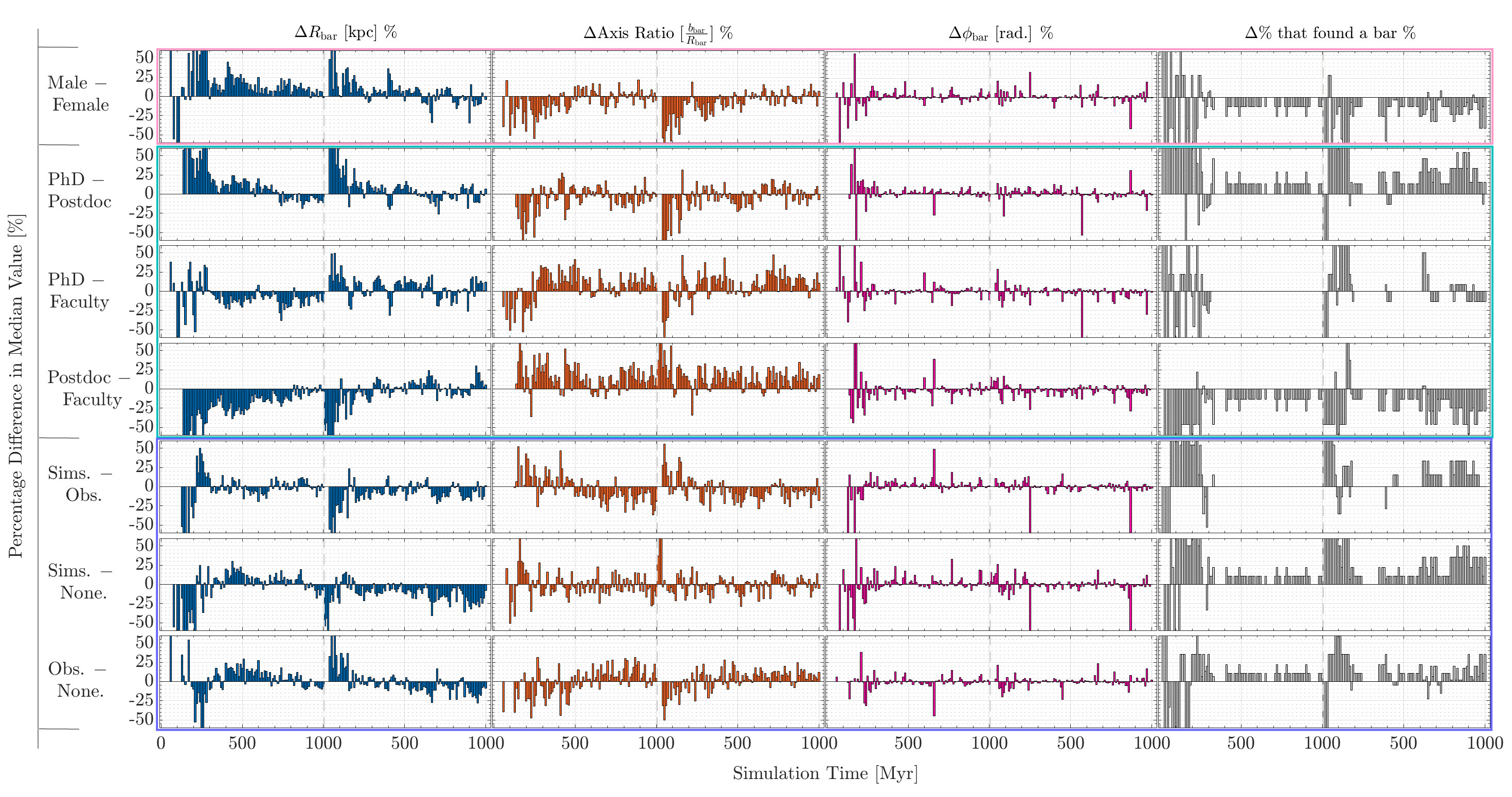}
    \caption{The percentage difference between the median value for the responses with each participant attribute in each snapshot for each parameter (columns): $R_{\rm bar}$, $b_{\rm bar}$, Axis Ratio, $\phi_{\rm bar}$, and fraction of individuals who found a bar at each time-step. A value of zero indicates no difference in the median between the two attribute distributions. Positive value indicates that the attribute listed first is greater than the attribute listed second, while a negative value is the opposite.}
    \label{f:percentdiff_barchar}
\end{figure*}

\subsection{The Impact of Astronomer Attributes on Classification}
Since a distinct range of responses in bar classification parameters was evident, we sought to assess whether this difference was easily attributable to some inherent feature of our population demographics and, as such, learn where to focus future studies to reduce these differences in our research outcomes. As an exercise, we consider the general bar parameters ($R_{\rm bar}$, Axis Ratio, $\phi_{\rm bar}$), as well as the fraction of participants who identified a bar as being present in a given snapshot, separated into the attributes of gender (male, female), academic career level (PhD, postdoc, faculty), and experience with studying bars in resolved galaxies (simulation, observation, none) as discussed previously in Section~\ref{ss:piecharts}. As indicated in earlier figures (Figure~\ref{f:piechart} \& Figure~\ref{f:barchart_postype}), these attributes are approximately evenly distributed individually throughout the full sample, as are all the other attribute types within each attribute group. To our surprise, we find that there are indeed some significant trends which arise when distributing the classification results in this manner. Although these can be considered small number statistics, we are representative of a community of astronomers contributing to research and the wider discourse with our findings. Any differences found between our classification, and possible correlations with the attributes we tested, may not exist within the larger population of practicing astronomers. On the strength of this test we simply cannot know, however, it does exist within our group which is a realistic example, or case study, of the way the field at large primarily produces works in these small, primarily institute-based, communities.

Figure~\ref{f:percentdiff_barchar} shows the percentage difference in the median results for each of the general parameters ($R_{\rm bar}$, Axis Ratio, $\phi_{\rm bar}$, and fraction of astronomers to identify a bar in a given snapshot). Here, the percentage difference is measured from the median value (see Figure~\ref{f:allbox}) of each attribute such that $(X_1-X_2)/0.5(X_1+X_2) \times 100$\%, where $X_1$ and $X_2$ are the medians of the two attributes being compared respectively. A positive value in this figure indicates that the attribute listed first on the $y-$axis ($X_1$) is greater than the attribute listed second ($X_2$) by the given percentage, while a negative value indicates the opposite is true. Naturally, a zero percentage difference value indicates that both distributions compared were sufficiently similar as to produce the same median response. 

Simply from a visual assessment of Figure~\ref{f:percentdiff_barchar}, we can notice the following trends. In terms of gender, male participating astronomers appear to consistently identify bars which have larger bar lengths ($R_{\rm bar}$). Female participating astronomers appear to consistently identify bars more often (in more snapshots) than their male counterparts and, in the early periods of bar growth ($\sim$1/2 simulation time) for each disk, as rounder (with larger axis ratios). Comparatively, both PhD students and faculty staff seem to identify bars with larger $R_{\rm bar}$ length than postdoctoral researchers for the most part. When comparing between each other, there is no clear trend overall, as PhD students find longer bars in the isolated disk (IsoB), while faculty appear to find longer bars in the tidal disk (TideB). On the other hand, the faculty staff find the thinnest, most ellipsoidal bars of all the career stages. Interestingly, postdocs seem the least optimistic about a bar existing at any given time period, with faculty and PhDs in general agreement for most periods once the bar has formed. If we consider, instead, how experience with resolved galaxies may affect the classification, it is evident that a similar number of time-periods have similar levels of percentage difference between all the sub-groupings of astronomers (simulation, observation, none) but any large-scale trends are less obvious. Simulators and observers, with experience studying bars in resolved galaxies, may be more optimistic about the presence of a bar compared with those astronomers who do not regularly deal with resolved galaxies in their day-to-day research. In the early periods of bar evolution for both disks, simulators appear to perceive rounder bars compared with observers and yet, in the later periods of bar evolution either the simulators become more conservative or the observers more optimistic as the bar grows, directly reversing the apparent trend in Axis Ratio. However, as these percentage differences vary more with bar evolution, it becomes difficult to make any kind of meaningful assessment of the trends or biases between the participant attributes simply by eye. 

\begin{figure*}[hbt!]
	\includegraphics[width=0.95\textwidth]{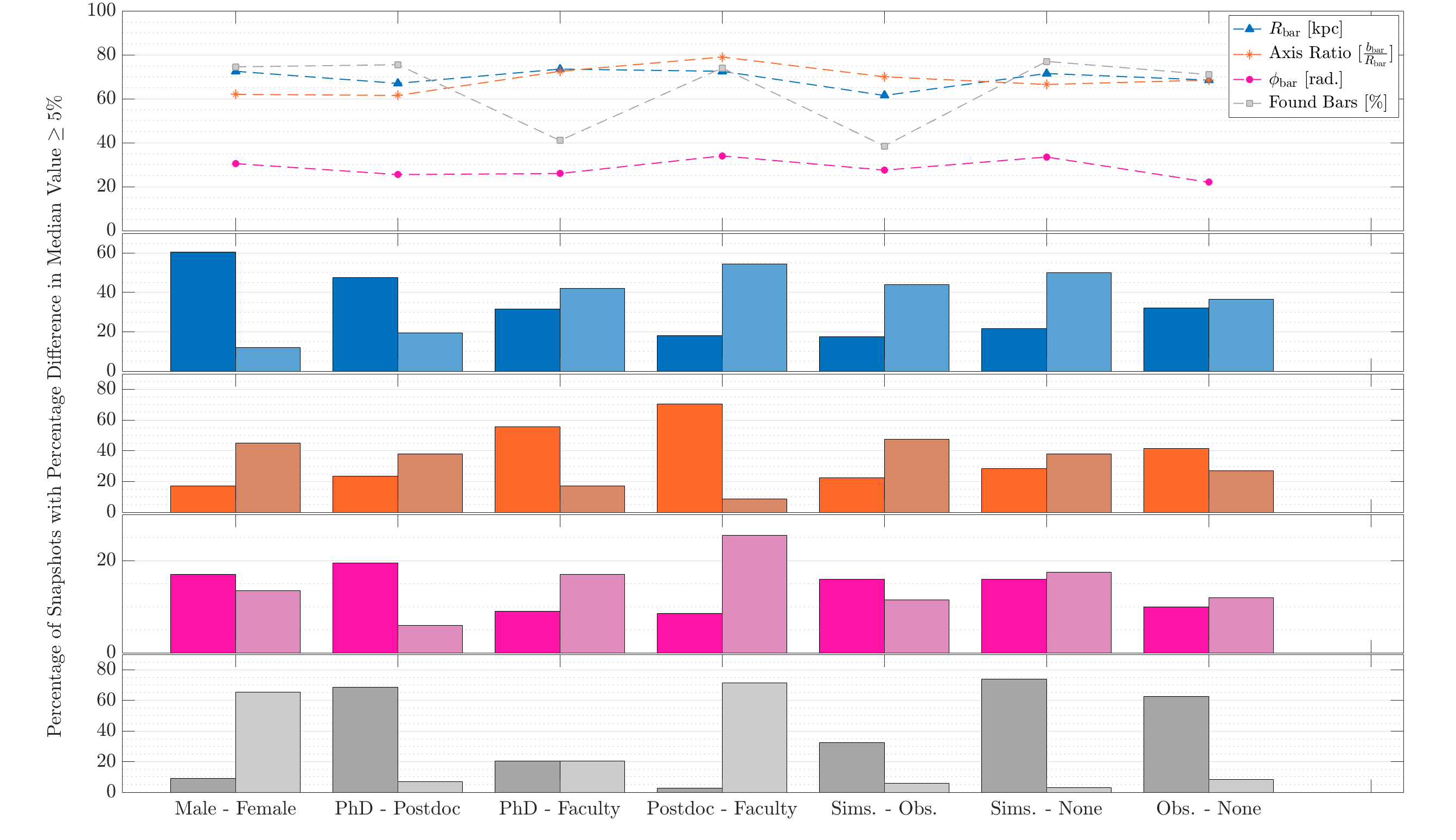}
    \caption{A measure of the fraction of snapshots with a percentage difference value higher than 5\% from Figure~\ref{f:percentdiff_barchar}. Top - the absolute difference for each of the four bar parameters: $R_{\rm bar}$ = blue triangle, Axis Ratio = orange star, $\phi_{\rm bar}$ = pink circle, \% to find a bar = grey square. Rows $2$--$5$ - the skew of values toward either the left or right group in the combination, as listed along $x-$axis coloured by the same parameter colour. Saturation only serves to better distinguish between left-right values.}
    \label{f:percentdiff_frac}
\end{figure*}

Here, we include Figure~\ref{f:percentdiff_frac} in an attempt to quantify these differences and any associated biases in a more direct representation. We use Figure~\ref{f:percentdiff_frac} to visually represent the relative significance of these differences arising between attributes of the contributing astronomers. The top panel of this figure displays the total fraction of snapshots where the median responses from the compared astronomer attributes (along the $x-$axis) differ by more than a $5$\% threshold value (introduced to account for systematic errors). Then, the subsequent four lower panels show the fractional contribution of each group to the total time spent in difference. That is, the higher the bar in the lower panels, the more time spent in difference (more bias), while a higher value in the upper panel simply indicates there is a large difference in the classification of that specific bar parameter between the two groups of astronomer attributes being compared. It is possible for the value in the upper panel to be large, signifying a large disagreement between astronomers, but for there to be no bias to either of the compared attributes (indicated by equally high bars in the lower panel), such as the $R_{\rm bar}$ (blue) measurement between observers and non-resolved galaxy astronomers (far right column). This would be an indication that this parameter is simply difficult to classify or that the inherent attributes being compared are unlikely to make significant impact to the result found.

If we purely consider the total difference between classification (upper panel), it appears that the largest difference is, in fact, in the definition of the Axis Ratio, as well as the bar length ($R_{\rm bar}$). The fraction to identify bars also seems to differ significantly, except between PhD students and faculty members, as well as simulators and observers of bars. The bar orientation angle ($\phi_{\rm bar}$) is clearly the parameter where most are in agreement, however, this too is not without some variation. In considering whether a bias exists within these differences, we compare the relative heights of the bars for each of the compared groups of astronomers aligned with the combination of attributes on the $x-$axis (different saturation simply distinguishes left and right groups for each combination). In this way, it is possible to assess by eye whether the total difference in the response for a given parameter is skewed towards one of the compared attributes. 

Some of the largest differences between bars in the lower panels of this figure are between the parameters specified by the participating male and female astronomers. These male astronomers, for example, find significantly larger values for the bar length ($R_{\rm bar}$) with percentage differences $\ge 5$\% for $\sim60$\% of galaxy snapshots, compared to their female counterparts, who find significantly larger bar lengths in only $\sim10$\% of galaxy images and most of these occur in the earliest time periods where it is not clear whether a bar even exists. Indeed, these female contributing astronomers are also significantly more optimistic about finding a bar to exist at all in a given time period ($\sim65$\% of the time periods) than the alternative, where more male astronomers believe bars to exist ($\sim10$\%). However, we also find a large difference is affected by the academic level of the astronomer. The postdoctoral researchers in our group of astronomers are much more likely to identify bars as more elliptical (large Axis Ratio: postdoc~$\sim70$\%, faculty~$\sim8$\%), less elongated (small $R_{\rm bar}$: postdoc~$\sim18$\%, faculty~$\sim55$\%) and less angled from the horizontal (small $\phi_{\rm bar}$: postdoc~$\sim8$\%, faculty~$\sim25$\%) than someone in a higher academic position (see central column). Additionally, they are significantly more pessimistic about a bar even existing at all when compared to both the academics and students on either side of academic age (small Found Bars: postdoc-faculty~$\sim2:72$\%, postdoc-PhD~$\sim7:68$\%). Interestingly, while the postdoc classifications of bars are very different from the faculty classifications, we do not see the same difference occurring between the academically younger population of PhD students and the faculty, particularly in their opinions about the existence (or non-existence) of bars in a given snapshot which is precisely equal ($\sim20$\%). Finally, we also find that astronomers who work with resolved galaxies (both simulators and observers) are much more optimistic about the presence of a bar than those who have no resolved galaxy experience (large Found Bars: sim-none~$\sim74:3$\%, obs-none~$\sim62:8$\%) but otherwise show little difference between each of the three experience groups in all classified parameters (max. between sims. and none groups on $R_{\rm bar}\sim28$\%).

\section{Significance}
\label{s:significance}
As there appears to be variation in the bar classifications of our group members, regardless of any or all attributes of our academic/personal identity which may be driving it, it follows that perhaps when we discuss bars as a discipline, at conferences, in a collaboration or even within a single research group such as ours, we may not be discussing the same features that we think we are. This has the potential to be a critical problem to our current research into bars in galaxies and here, we demonstrate with a simple example.

\subsection{Bar Length \& Star Formation Profiles}
One of the most significant differences in classification which appeared during the \emph{findAbar} process was in the extent of the bar length ($R_{\rm bar}$). This varied across the whole set of astronomer attributes but was most prominent between male and female identifying astronomers, as well as between postdocs and faculty. We can extrapolate that if such a bias exists and is prevalent within the entire community, then we must begin to question the impact of such classification variance on the science we produce. Here, this is demonstrated with the distribution of SFR and SFE across the bar in galaxies which is an area of some contention within the field as it stands. While it is well accepted that bars have significant impact on the star forming properties of their host galaxies \citep[e.g.][]{Downes1996, Sheth2002, Momose2010, Watanabe2019, Querejeta2021, iles2022}, studies have demonstrated that while the bar generally enhances the star formation in the central regions of their host galaxies, how effectively gas is converted to stars in the bar region may be more nuanced. There are studies which show that the SFE is consistently lower in the bar region \citep[e.g.][]{Momose2010, Watanabe2011, Muraoka2016, Pan2017}. There are also studies which show that the SFE is not systematically lower in the bar region \citep[e.g.][]{Muraoka2019, Diaz-Garcia2021, Querejeta2021}. Then, to add further complication, there are also studies which show that the SFE may be lower within the bar region but it is in fact enhanced at the bar-ends \citep[e.g.][]{Handa1991, Hirota2014, Law2018, Yajima2019, Watanabe2019, Maeda2020}.

\begin{table}
    \centering
    \begin{tabularx}{\textwidth}{c|c|c|c}
        \textbf{Galaxy ID} & \textbf{Morphology} & \begin{tabular}[c]{@{}c@{}}\textbf{\citet{maeda2023}}\\\textbf{SFE Profile Shape}\end{tabular} & \begin{tabular}[c]{@{}c@{}}\textbf{This Work }\\\textbf{SFE Profile Shape}\end{tabular}\\
        \hline
        NGC\,1300 & SBbc & Peak & Flat\\
        NGC\,1365 & SBb & Peak & Peak\\
        NGC\,2903 & SABbc & Peak & Peak\\
        NGC\,3627 & SABb & Peak & Peak\\
        NGC\,4303 & SABbc & Peak & Flat\\
        NGC\,4535 & SABc & Fall & Fall\\
        NGC\,4548 & SBb & Peak & Peak\\
        NGC\,4579 & SABb & Peak & Peak\\
        NGC\,5236 & SABc & Peak & Flat\\
    \end{tabularx}
    \caption{The nine example galaxies selected from \citet{maeda2023} are listed with the galaxy identifier (left), morphology from the Third Reference Catalogue of Bright Galaxies \citep[RC3; ][]{DeVau1991} and the shape of the SFE profile within the limit of the bar radius $R/R_{\rm bar}=1$ as defined from \citet{maeda2023} and the this work (right).}
    \label{t:maeda_compare}
\end{table}

It is at this point that we encourage the reader to stop and reflect on their definition of what might constitute a \textquoteleft bar-end’ region within a given barred galaxy. It is possible to extract such a bar-end region from the population belonging to the bar component; the arm component; or, some component partially overlapping both bar and arms with equal or non-equal contributions from each. It is also not common to specify precisely where this region has been sampled within the galaxy, even if the bar classification has been well documented. Regardless, it follows that with a variable classification of the bar extent ($R_{\rm bar}$) which may be dependent on the specific attributes of the astronomer doing the classification, the precise regions of a given galaxy which constitute \textquoteleft bar’, \textquoteleft not-bar’ and even the \textquoteleft bar-end’ may not, in fact, be representative of the same morphological component to be compared. 

\begin{figure*}[hbt!]
	\includegraphics[width=\textwidth]{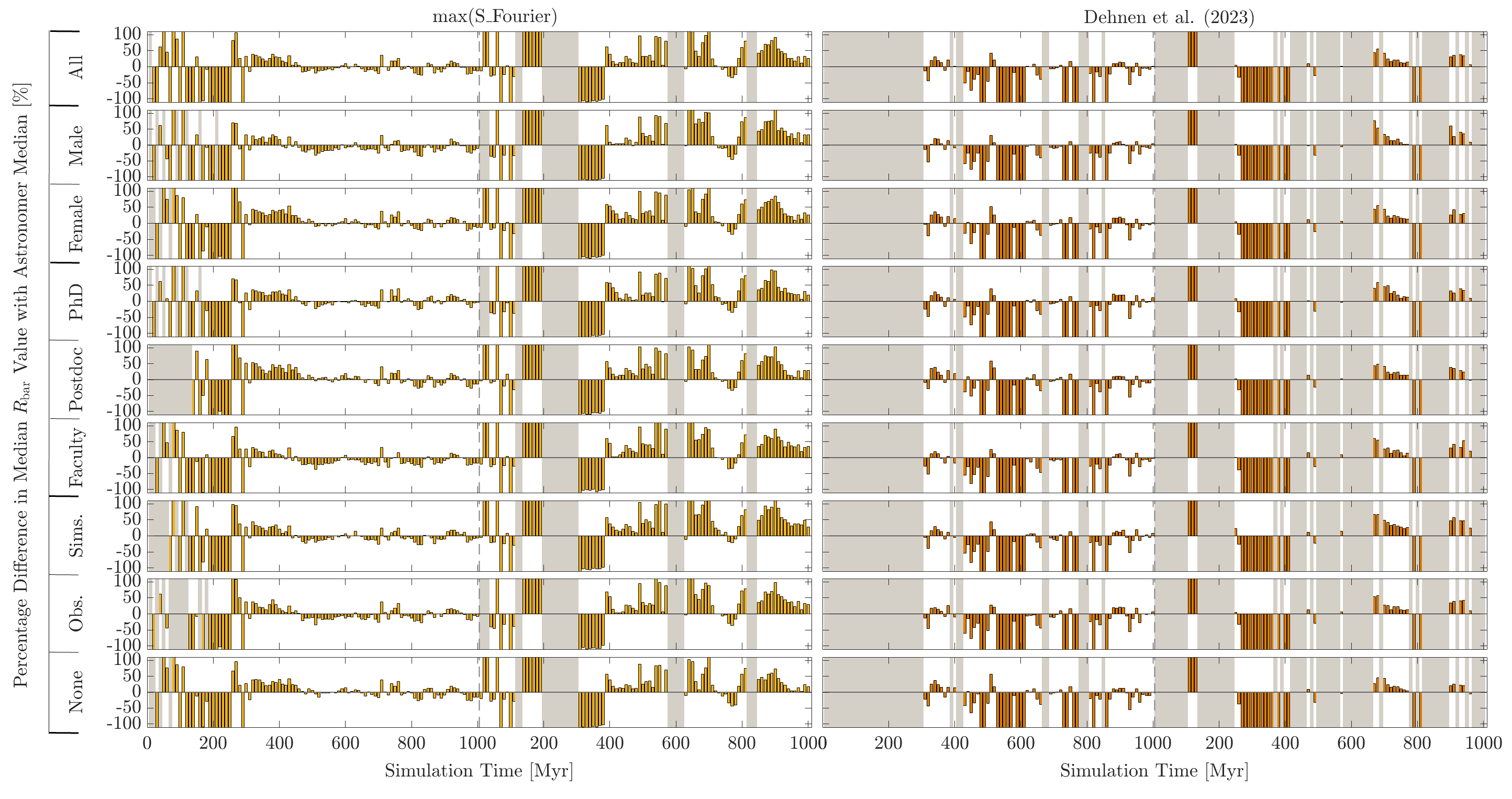}
    \caption{The percentage difference between the median value for $R_{\rm bar}$ in all the responses (top row) and with each participant attribute (lower rows), compared between the two automated methods (columns). A value of zero indicates no difference in the median between the two attribute distributions. A positive value indicates that the value from the automated method is greater than the attribute listed, while a negative value is the opposite. Grey regions indicate time periods where there was no value recorded for at least one of the responses compared.}
    \label{f:percentdiff_lit_Rbar}
\end{figure*} 

We demonstrate this using the SFE values of 10 barred galaxies with different morphological types \citep[i.e. SABb, SABbc, SBb etc. as per][]{DeVau1991} selected from the work of \citet{maeda2023}. These galaxies were all selected from the PHANGS-ALMA sample \citep{leroy2021} and the SFE profiles determined by the SFR from FUV (far ultraviolet) and WISE (Wide-field Infrared Survey) data combined with a molecular gas density from the CO(1-0) or CO(2-1) moments \citep{maeda2023}. In \citet{maeda2023}, these profiles are presented relative to a $R/R_{\rm bar}$ length scale with these $R_{\rm bar}$ values from \citet{Herrera-Endoqui2015}. We used the same original galaxy data product used by \citet{Herrera-Endoqui2015} to define the bar, then de-projected these galaxies and calculated our own $R_{\rm bar}$ by eye with the \emph{findAbar} method. This was, however, only completed once per galaxy by one author and then corrected to the median values from \emph{findAbar}. This is not the most rigorous approach possible but reasonably serves to demonstrate the impact of $R_{\rm bar}$ variation. Finally, these values were used to compare the shape of the profile within the bar region with the results of \citet{maeda2023}. Table~\ref{t:maeda_compare} is a summary of this comparison for 9/10 sample galaxies, as NGC\,4536 was originally selected from \citet{maeda2023} but subsequently omitted due to low pixel numbers with $\Sigma_{\rm mol}\ge5$M$_\odot$pc$^{-2}$). Here, the SFE profile is described by shape; either \textquoteleft Peak’ for a profile that peaks in the centre and at the edge where $R/R_{\rm bar}=1$, \textquoteleft Fall’ for a profile that peaks in the centre and continues to decrease towards the edge where $R/R_{\rm bar}=1$, or \textquoteleft Flat’ if there is no obvious peak at the $R/R_{\rm bar}=1$ edge of the profile or decrease in SFE over the bar length. 

The results in this table simply serve to demonstrate that it is possible to observe different trends in SFE across the bar if the bar classification is completed by a different method or different individual. With only nine example galaxies, arbitrarily selected, we find at least 3/9 galaxies where a difference in shape of the SFE trend within the bar would be reported. A further 3/9 galaxies also had significant differences between the two classifications (total 6/9 with large $\Delta R_{\rm bar}$) but, despite these differences, we would have still reported similar shapes of the SFE for these three galaxies. Furthermore, on the galaxies where we differed in bar extent ($R_{\rm bar}$) classification, it was mostly to classify a shorter bar, with only one galaxy where we identify a longer bar (and no difference in the SFE profile shape) compared to \citet{maeda2023}. Despite being a trivial example, this is a large percentage ($\sim33$\%; or 66\%) of cases where the difference in classification would make a significant difference to any scientific results reported. 

\section{The Solution is...}
\label{s:solutions}
Thus far, we have simply highlighted the differences in classifications of bars by our group of astronomers and demonstrated simply the significance that such differences might have on our current and future understanding of these galaxies. However, even if we know such a problem exists now, what are we to do about it?

\subsection{Computers?}

One solution is to use an automated, ideally non-parametric, bar-finding algorithm or program. The difficulty with this is that among the current popular (publicly available) methods, certain methods work better for certain types of data or have a high knowledge threshold, if they are to be used independently of assistance from the developer. If a method is not comparable for all research circumstances, is the result any more functional than a biased definition from a human astronomer? In that case, at the very least, the lead author of the paper can be contacted to explain their definition, or to provide further assistance in the process of comparison. Understanding the biases behind an automated method can be much more obtuse, often hidden behind fundamental computing processes, and including the likelihood of unconscious operator error.

As a demonstration, we determined the bar extent ($R_{\rm bar}$) via two computational methods which take advantage of the Fourier modes to define a bar for the same simulated snapshots used in \emph{findAbar} and compare these with our own median classification (see Figure~\ref{f:percentdiff_lit_Rbar}). Here, we use a recently popular code for finding the barred region and pattern speed in (primarily simulated) galaxies, which is publicly available and regularly updated, as of the time of writing \citep[see ][]{dehnen2023}. Additionally, we also use a more straightforward algorithm but similar method that also makes use of the a strength parameter ($S={\rm rms}(A_{m={\rm even}})-{\rm rms}(A_{m={\rm odd}})$), where $A$ is the amplitude of the Fourier mode and $A_{m={\rm even}} = A_2, A_4, A_6$ while $A_{m={\rm odd}} = A_1, A_3, A_5$ from the \textsc{pynbody} Fourier profiles). A naive extent for the bar is set at radius where this $S$ value is maximum. Figure~\ref{f:percentdiff_lit_Rbar} is similar to the previous Figure~\ref{f:percentdiff_barchar}, however, here we only focus on the one parameter: the bar length ($R_{\rm bar}$) and compare the astronomer responses in the total sample (top row), as well as different attribute groups (lower rows), with the two automated methods (right: \citet{dehnen2023}; left: max $S$). Again, a zero value means there is no difference between the bar length found in the automated method and the median of our individually classified responses. A positive value will indicate that the automated method produced a larger value of $R_{\rm bar}$, while a negative value will mean the opposite is true. For the periods where there was no value recorded for at least one of the responses, a grey shading effect is used to block out the unresolved region for each comparison. This may be due to either the situation where the number of astronomers who found a bar in that period were too few to calculate a median or that the automated method simply did not identify a bar in that snapshot.  

The most obvious feature of this comparison is that there are many missing periods, especially in the tidal disk (TideB; right). The max $S$ method appears to find bars in more snapshots than the method of \citet{dehnen2023}, however, most of these periods have differences of more than $50$--$100$\% and probably should be considered as possible misidentification rather than true bar finding. While the maximum $S$ method does appear to find reasonably comparable values for most of the periods in the isolated disk after $\sim 400$\,Myr (IsoB; left), it also seems to struggle with classifying the bar in the tidal disk. The method of \citet{dehnen2023} also has particularly high numbers of snapshots with very large percentage differences. This is due to this routine preferentially finding very small bars ($\le0.05$\,kpc). We note, however, that this may not be the ideal application of this method, and acknowledge that operator error may somewhat have contributed to these results, as well as the presence of a forming nuclear stellar disk in the centre of this galaxy. For the periods where a somewhat comparable bar length was identified by these automated processes, no obvious bias to consistently larger or smaller bars than those identified by all possible astronomer subsets is evident, at least for the isolated disk, which seems promising for the future application of these types of algorithms. 

However, the relative inconsistency of these results serve only to highlight how difficult it is to be confident in the current automated methods used for identifying the bar, especially where the automated bar lengths differ considerably from the similar bar parameters defined by eye. We, therefore, raise the question whether this is a problem in our relatively simple automation methods or our combined understanding of what constitutes a bar overall. 

An alternative solution many have suggested is the recently popular machine learning and AI. However, these computational methods must currently be trained by some given set of rules or on some kind of existing \textquoteleft known’ data-set. Both of these training methods are inevitably subject to the opinions (and inherent biases) of the researcher behind the development of the tool or method. This is the same problem for any open crowdsourcing attempt to classify morphological features. For a simple algorithm or visual classification, if we were unsure of how a bar in a given study was defined but wished to compare with it, we could reach out to the corresponding author of the study directly. This added level of complexity in defining the bar for the machine learning program or survey participants to learn \emph{from}, means that details of the original classification method can easily disappear into obscurity. This will make it increasingly difficult to identify any inherent biases in these methods and our belief that we are comparing like-with-like must increasingly come from faith. Unless we, as a field, can decide on one toolset, one theoretical definition and/or the corresponding best practice for identifying the barred region in galaxies, these results indicate that we should proceed with caution when developing in this area. Otherwise, the number of inherent and indeterminate biases will increasingly become a source of error for any general or universal statements regarding the features or processes of bars in galaxies. 

\subsection{An Interim Best Practice}
If computers are not currently a practical solution, what can we do to resolve this issue? Of course, the ideal solution would be a broad, thorough and widely applicable definition of the bar. Or, alternatively, a standard mechanism similar to the effective radius (or half-light radius) of galaxies which we can use to practically describe the bar extent. The development of a (digital) toolset which works to define the bar regardless of the data product and the bar strength, which everyone in the community agrees to use, would also go a long way to a solution. However, these are all long term goals which one may spend an entire academic career working on without a viable solution. 

\begin{figure}[hbt!]
\includegraphics[width=\columnwidth]{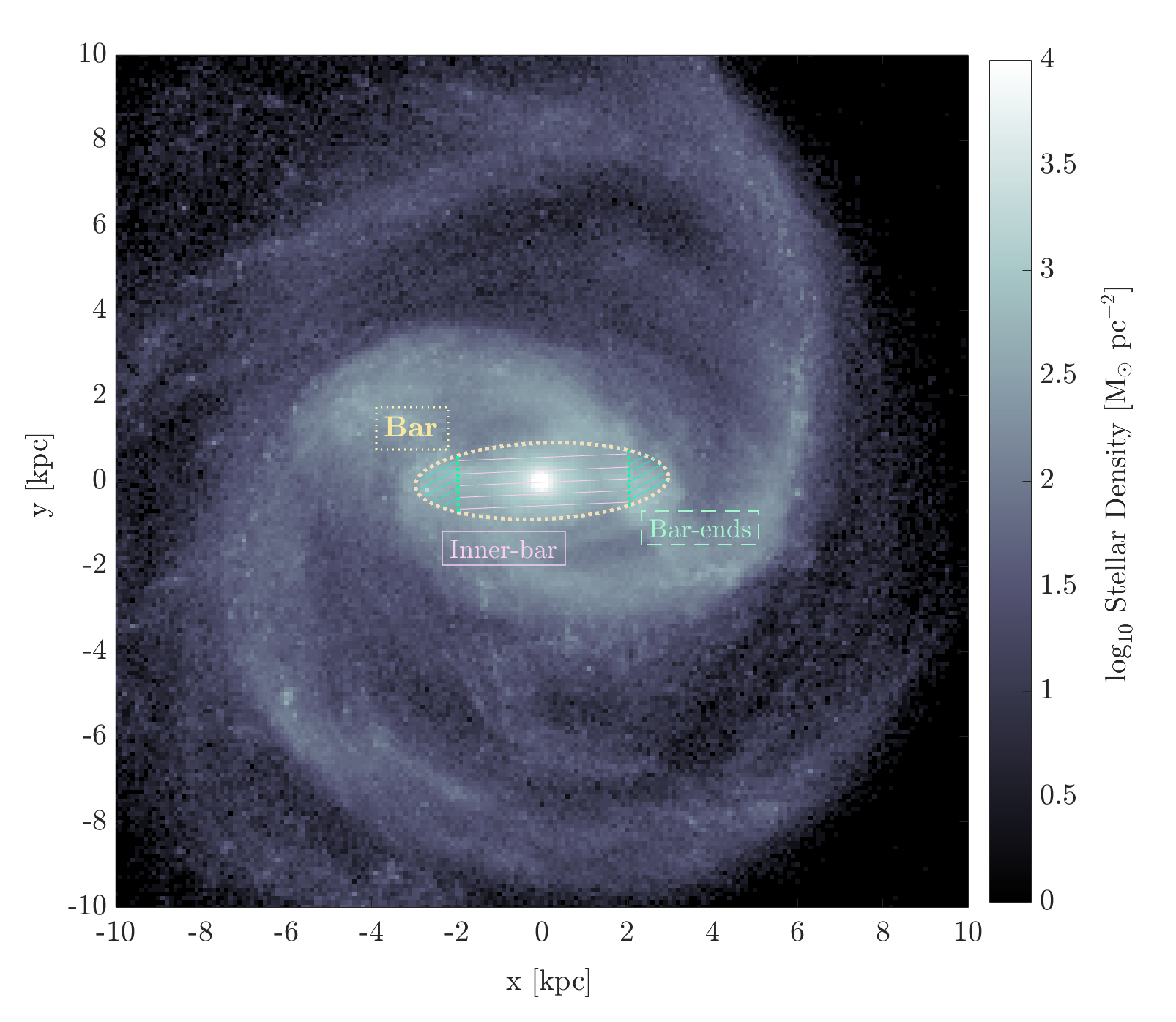}
    \caption{A standard face-on stellar density map for one of the barred galaxies included in \emph{findAbar} \citep[IsoB; ][]{iles2022, iles2024} with the bar region highlighted by an ellipse and bar-ends accounting for the outer 5\% of the bar length hashed. The inner-bar region is also identified with text and stripes, to differentiate this region from the full bar region which would also include the bar-end components.}
    \label{f:example}
\end{figure}

So, in the interim, the only and best solution must be to be more articulate about the bars we are defining and how we use them. This may sound trivial and many would argue this is already occurring. However, to what extent it is already occurring and how mindful we are of new members to the field when articulating these classifications is something we must always be actively considering when we present our research. If possible, we would like to recommend every member of the community to include an image, either within the article, as appendices, or even in some supplementary material storage, which clearly demonstrates all the regions discussed and analysed. A table or discussion of parameters, including bar extent, pitch angle, axis ratio and whether the bar is mapped by an ellipsoid or rectangular region would be the minimum requirement recommended. Additionally, if making reference to the \textquoteleft bar-end’ region of the galaxy, it is important to note whether this region has been taken from the binary \textquoteleft bar’ or \textquoteleft arm’ component or both, and ideally to also document this in the image. An example of this kind of image classification is included in Figure~\ref{f:example} using the IsoB galaxy image from Figure~\ref{f:matlab_bars} and the median \emph{findAbar} responses with a bar-end accounting for 5\% of the original bar region. 

\section{Conclusions}
\label{s:conclusion}
The original intention behind this research was simply to ascertain whether comparisons between simulated galaxies with bars could be considered consistent with observations of real barred galaxies. Initially, we sought to investigate this by comparing how bars in one simple galaxy image would be classified by the astronomers who complete these simulations and observations, as well as a control sample of astronomers who should be able to identify a bar from their elementary astronomy understanding but do not work with such galaxies in their day-to-day research. Thus, our $21$ astronomers with varied experience in studying resolved galaxies, with also differing career stages and gender, came together to assess $200$ evolutionary snapshots, spanning the early phase of bar evolution, in two differently barred galaxies from simulation. This process has served as a test of the quality and consistency of our \textquoteleft expert’ classification of the bar in galaxies on the scale of a standard astronomy department currently undertaking research in this area.  

Analysing these $4\,000+$ classifications, we found that it is indeed possible for the bars identified by our range of different astronomers to differ significantly in all the standard bar classification parameters: bar length, axis-ratio, pitch-angle and even whether a bar is present at all. Additionally, this may not be entirely due to some kind of random scatter. The gender and academic level of the astronomers identifying a given bar is likely to contain a significant bias, particularly in the definition of bar length ($R_{\rm bar}$) and size (Axis Ratio) parameters. It is currently unclear whether this career level dependence can be specifically related to the experience of these astronomers with analysing bar features in resolved galaxies in particular (in either simulated or real, observed data) or may simply be related to exposure to astronomy as a field. The possible existence of such inherent bias in our understanding of the bar feature is concerning, as such difference has implications for interpreting morphological trends, such as bar-end effects, which may be significantly dependent on the bar length classification ($R_{\rm bar}$).

In summary, we list here four main features we observe in the analysis of the \emph{findAbar} dataset for our group:
\begin{itemize}
    \item The most reliable property of a bar to measure is the orientation (position angle, $\phi_{\rm bar}$).
    \item Simulators and observers who work with bars seem to be in agreement of bar properties and are more optimistic about the presence of a bar than non-specialists. 
    \item Male and female astronomers in particular are classifying the same bars as extending to different disk radii (bar length, $R_{\rm bar}$).
    \item There may be some kind of paradigm shift in classifying the same bars which may correlate with academic level. Despite current PhD students appearing to resonate with faculty rather than the closer-in-age postdoctoral researchers, it remains unknown if these students will continue to do so throughout their careers. 
\end{itemize}

Through the example of bar-related SFE, by focusing on the third point (the difference in $R_{\rm bar}$ between male and female astronomers), we have aimed to highlight how these differences might affect a scientific result. In this example, we show that while the trend of SFE being suppressed within the bar region, and possibly enhanced at the bar-ends of some galaxies but not others, cannot be immediately explained by a case of female observers classifying shorter bars on average, the impact of such a bias on the existing results, equally, cannot be immediately discounted. Additionally, this is only one trivial example from the literature. On the strength of this, we posit many more cases should similarly exist and be subjected to some scrutiny in the future by those working in this area. 

Finally, there is no clear solution at this time. Automation does not currently seem to be a viable option based on the bar-finding algorithm methods tested. Both the Fourier-based algorithms either failed to identify a bar in snapshots where most astronomers agreed one should exist or produced bar length ($R_{\rm bar}$) values which were over $50$--$100$\% different to the median bar lengths we determined. However, for the few snapshots where there was some reasonable agreement, these algorithms appeared to demonstrate no obvious bias to longer or shorter bar lengths, so perhaps there is hope for this solution yet. Machine learning, AI and crowdsourcing are also regularly proposed as a solution, but we caution against possible human biases being obscured or even completely hidden within the training routines of these tools. 

The ideal solution is that the community at large should come to a consensus on the theoretical definition of a bar, develop a standard measure for the bar extent (as the effective radius is to galaxy disk limits), or produce a tool which is able to classify bars in a variety of data products, with varying bar properties and in a manner which is consistent with the visual expectations of astronomers. However, we also recognise that this may be an insurmountable goal under a reasonable time-frame, so we advocate for more thorough articulation of the bar classification parameters to become a community best practice, including diagrammatic representations where possible. 

Thus, we list here our goals for the future of bar classification:
\begin{itemize}
    \item Everyone in the community to be more clear in describing how they have defined the \textquoteleft bar’ region (and especially the \textquoteleft bar-end’ region) with full lists of parameters and diagrams available when presenting their research. 
    \item The field at large to come to an agreement on a broadly applicable definition for the bar in theory, practice and expression. 
  \item A more comprehensive study with a larger sample of astronomers and galaxy images to better characterise the unconscious bias in our \textquoteleft expert’ classification of bars.
\end{itemize}

Bars are important to many aspects of galaxy research and, it is only by understanding how we individually may perceive these features, that we can hope to piece together a complete understanding of their formation, evolution and impact on the wider universe in the future.

\begin{acknowledgement}
Images and partial analysis were made using the \textsc{pynbody} (\url{https://github.com/pynbody/pynbody}) \textsc{python} package of \citet{Pontzen2013} and the \citet{matlab} \textsc{matlab} programming and numeric computing platform. This research has made use of the Astrophysics Data System, funded by NASA under Cooperative Agreement 80NSSC21M00561. Observational data was accessed from the PHANGS-HST survey \citep{Lee2022}. Numerical computations were carried out on the Cray XC50 at Center for Computational Astrophysics, National Astronomical Observatory of Japan. Thanks to Thorsten Tepper-Garcia for their input to this project. Finally, many thanks to the anonymous referee for their very helpful comments and suggestions, as well as to the editors who have worked hard to facilitate the publication of this manuscript. 
\end{acknowledgement}

\paragraph{Funding Statement}
CC acknowledges the support the Australian Research Council (ARC) through the Discovery Projects DP190100666 and DP210103119. 
EFK acknowledges the support provided by an Australian Government Research Training Program (RTP) Scholarship.
EJI acknowledges the support of the Australian Research Council through Discovery Project DP220103384. 
ELS was supported by the Australian Government through the Australian Research Council Centre of Excellence for Dark Matter Particle Physics (CDM, CE200100008).
MIL acknowledges the support from Comunidad de Madrid in Spain (grant Atracci\'{o}n de Talento Contract no. 2023-5A/TIC-28943).
KR thanks the LSST-DA Data Science Fellowship Program, which is funded by LSST-DA, the Brinson Foundation, and the Moore Foundation; Their participation in the program has benefited this work. 
YM is supported by an Australian Government Research Training Program (RTP) Scholarship.
ST acknowledges the support from the Royal Thai Government Scholarship and the University of Sydney Postgraduate Research Supplementary Scholarship in Understanding the Milky Way through Integral Field Spectroscopy.
MGP acknowledges the support by the Professor Harry Messel Research Fellowship in Physics Endowment, at the University of Sydney.

\paragraph{Competing Interests}
None.

\paragraph{Data Availability Statement}
The data underlying this article can be shared upon reasonable request to the corresponding author. As of publication, we are currently developing a web-based, open access version of \emph{findAbar} which will be more rigorous and have significantly more depth. We encourage the community to participate in this study in the future. Please feel free to contact the corresponding author if you are interested in the current state of development for \emph{findAbar2.0}.

\bibliography{mybib.bib}

\end{document}